\documentclass[prd,aps,showpacs,reprint,preprintnumbers,amsmath,amssymb,superscriptaddress,longbibliography,nofootinbib,floatfix]{revtex4-2}

\usepackage[utf8]{inputenc}
\usepackage{amsmath}
\usepackage{amssymb} 
\usepackage{enumerate}
\usepackage{multirow}
\usepackage{braket}
\usepackage[toc,page]{appendix}
\usepackage{url}
\usepackage{slashed}
\usepackage{graphicx}
\usepackage[pdftex,bookmarks,linktocpage,pdfpagelabels,plainpages=false,hyperfigures,linkcolor=blue,citecolor=blue]{hyperref} 
\hypersetup{colorlinks=true}
\usepackage{adjustbox}
\usepackage{mathrsfs,amssymb}  
\usepackage{cancel}
\usepackage{xcolor}
\usepackage[normalem]{ulem}
\usepackage{cleveref}
\usepackage{tikz}
\usetikzlibrary{decorations.markings}
\usetikzlibrary{positioning}
\usetikzlibrary{shapes}
\usetikzlibrary{calc}
\usepackage{tikz}
\usepackage{tikz-feynman}
\newcommand{\beq}{\begin{equation}}
\newcommand{\eeq}{\end{equation}}

\newcommand{\bea}{\begin{eqnarray}}
\newcommand{\eea}{\end{eqnarray}}

\usepackage{tikz-feynman}
\usepackage{amsmath}
\tikzfeynmanset{compat=1.1.0}
\usepackage{comment}
\begin{document}

\title{Predicting Three Generations of Fermions: Discovery Prospects of the Bilepton Model}
\author{Andreas Crivellin}
\email{andreas.crivellin@cern.ch}
\affiliation{Universitat Autònoma de Barcelona, 08193 Bellaterra, Barcelona.}
\affiliation{ICREA, Instituci\'o Catalana de Recerca i Estudis Avan\c{c}ats,
Passeig de Llu\'{\i}s Companys 23, 08010 Barcelona, Spain.}

\author{Paul H.~Frampton}
\email{paul.h.frampton@gmail.com}
\affiliation{Dipartimento di Matematica e Fisica "Ennio De Giorgi",
Universit\`{a}  del Salento and \\ INFN-Lecce, Via Arnesano, 73100 Lecce, Italy.}

\author{Ahmed.~Hammad}
\email{hammad@kias.re.kr}
\affiliation{Center of AI and Natural Sciences, KIAS, Seoul 02455, Korea}
\affiliation{Theory Center, IPNS, KEK, 1-1 Oho, Tsukuba, Ibaraki 305-0801, Japan.}

\begin{abstract}
We study the production of pairs of doubly-charged bileptons and assess their discovery potential in light of the integrated luminosities available at the High-Luminosity LHC. The production rates are governed primarily by the bilepton mass, $(m_Y)$, and the mass of the exotic heavy quarks, $(m_D)$. We consider two complementary production channels: (i) direct bilepton pair production and (ii) bilepton production mediated through heavy-quark decays. Notably, the latter typically yields significantly enhanced cross sections and gives rise to distinctive LHC signatures, even when the bileptons are produced off-shell. The two mechanisms therefore probe complementary regions of parameter space, with direct production being predominantly sensitive to $m_Y$, while the heavy-quark-mediated channel depends mainly on $m_D$. Owing to the essentially background-free signature of four energetic leptons at the LHC, we show that Run-2 data allow a discovery only for $m_D \lesssim 1\,\mathrm{TeV}$, whereas the HL-LHC can achieve a $5\sigma$ discovery up to $m_D \lesssim 2.5\,\mathrm{TeV}$ (nearly independently of $m_Y$) and/or for $m_Y \lesssim 2\,\mathrm{TeV}$ (even if $D$ is heavy).
\end{abstract}
\maketitle

\noindent
\section{Introduction}

The Standard Model (SM) of electroweak interactions~\cite{Glashow:1961tr}, dating back to the 1960s, has survived all experimental attempts~\cite{ParticleDataGroup:2024cfk} to find a definitive chink in its armour (disregarding the evidence for neutrino masses). It is based on the gauge group $SU(2)_L \times U(1)_Y$, as the smallest group which can accommodate charged $W^{\pm}$ bosons and a massless photon. It predicted a neutral current which was observed indirectly at CERN in 1973~\cite{GargamelleNeutrino:1973jyy}, while the corresponding massive $Z$ boson was discovered, together with the $W$ boson, in 1983~\cite{UA1:1983crd,UA2:1983tsx,UA1:1983mne,UA2:1983mlz} (also at CERN).

On the theoretical side, incorporating the electroweak description into a consistent renormalisable field theory required the introduction of spontaneous symmetry breaking through the Brout-Englert-Higgs mechanism~\cite{Higgs:1964ia, Englert:1964et}. The specific choice of a single $SU(2)_L$ doublet scalar was made in Ref.~\cite{Weinberg:1967tq}, and the accommodation of quarks was
provided in Ref.~\cite{Glashow:1970gm}. Importantly, anomaly cancellation is achieved within each complete generation of quarks and leptons, which allowed Ref.~\cite{Glashow:1970gm} to predict the charm quark. The particles in the SM were completed by the top quark discovery in 1995~\cite{CDF:1995wbb,D0:1995jca} and of the Brout-Englert-Higgs (BEH) boson~\cite{ATLAS:2012yve,CMS:2012qbp} in 2012. While the BEH boson remains the only particle discovery made so far at CERN's LHC, this does not exclude a future discovery, in particular in light of the forthcoming High-Luminosity (HL) LHC.

With the established three quark families, together with matching leptons and massive neutrinos, the SM has at least twenty-eight parameters (12 fermion masses, 10 mixing angles and phases (CKM and PMNS), 3 gauge couplings, the Higgs mass, the $Z$ boson mass and the $\bar{\theta}$ term, none of which even today has a convincing theoretical derivation. 

Here we want to focus on the number of generations (${\cal N}_f$).Ref.~\cite{Kobayashi:1973fv}
which showed that explicit $CP$ violation in the weak interactions
requires at least three families, but this was merely a lower bound
${\cal N}_f \geq 3$. However, when Glashow wrote his most famous paper~\cite{Glashow:1961tr} in 1960, quarks were unknown, and even the triangle anomaly's importance was not evident. If Glashow knew all of these things in 1960, it is hypothetically possible that he would have gone beyond the $SU(2)_L \times U(1)$, which treats each family as identical? In this regard, it is interesting that both Weinberg~\cite{Lee:1977qs} in 1977 and Glashow~\cite{Glashow:1984gc} in 1984 considered using $SU(3)_L$ as a part of the electroweak gauge group, but not with a view to deriving the value of ${\cal N}_f$.

It was more than three decades after 1960 before the bilepton model emerged in 1992 with electroweak gauge group $SU(3)_L \times U(1)$ and with inter-family anomaly cancellation, which requires ${\cal N}_f$ to be a multiple of three~\cite{Frampton:1992wt}.\footnote{Ref.~\cite{Pisano:1992bxx} presented a model very similar to Ref.~\cite{Frampton:1992wt} but did contain one error and one omission. It must be the third family, which is treated differently, with the first two being identical, to maintain the GIM mechanism and remain phenomenologically viable with respect to weak neutral interactions. The omission concerns the upper energy limit on the symmetry breaking, which requires the new physics to be at the TeV scale.} Insistence on the necessary asymptotic freedom of QCD then allows only ${\cal N}_f \equiv 3$ exactly.\footnote{In fact, the combination of precision observables and LHC searches now excludes a chiral fourth generation~\cite{Djouadi:2012ae,Eberhardt:2012sb}.} Thus, if the bilepton model is adopted by Nature, one of the 29 SM parameters has a compelling theoretical derivation. 

\section{Model}

We consider the original bilepton model introduced in Refs.~\cite{Frampton:1992wt}, with the gauge symmetry
\begin{equation}
  \mathcal{G} = SU(3)_c \times SU(3)_L \times U(1)_X\,,
\end{equation}
where the electric charge operator is defined as
\begin{equation}
  Q = T_3 + \sqrt{3}\,T_8 + X\,.
  \label{eq:charge}
\end{equation}
Here $T_3$ and $T_8$ are the diagonal generators of $SU(3)_L$ and $X$ denotes the $U(1)_X$ charge.

The fermion sector is arranged non-universally under $SU(3)_L$. The first two quark families are triplets,
\begin{equation}
  Q_{1} =
  \begin{pmatrix} u_L \\ d_L \\ D_L \end{pmatrix},\quad
  Q_{2} =
  \begin{pmatrix} c_L \\ s_L \\ S_L \end{pmatrix},\quad
  Q_{1,2}\in\!\left(3,3,-\tfrac{1}{3}\right),
  \label{eq:Q12}
\end{equation}
while the third family transforms as an anti-triplet,
\begin{equation}
  Q_{3} =
  \begin{pmatrix} b_L \\ t_L \\ T_L \end{pmatrix},\quad
Q_{3}\in\!\left(3,\bar{3},+\tfrac{2}{3}\right)\,.
  \label{eq:Q3}
\end{equation}
The states $D$, $S$, and $T$ are new TeV scale quarks with electric charges
$-4/3$, $-4/3$, and $+5/3$, respectively. The right-handed quark
singlets transform as $(d_R,s_R,b_R)\in(\bar{3},1,+\frac{1}{3})$,
$(u_R,c_R,t_R)\in(\bar{3},1,-\frac{2}{3})$,
$(D_R,S_R)\in(\bar{3},1,+\frac{4}{3})$, and
$T_R\in(\bar{3},1,-\frac{5}{3})$.
Cancellation of the $SU(3)_L^3$ gauge anomaly requires all three
lepton generations to transform as anti-triplets,
\begin{equation}
  \ell_f =
  \begin{pmatrix} \ell^-_L \\ \nu_{\ell L} \\ \ell^c_R \end{pmatrix},
  \quad
  \ell_f \in (1,\bar{3},0),\quad f = e,\mu,\tau,
  \label{eq:leptons}
\end{equation}
where the superscript $c$ stands for charge conjugation.

The $SU(3)_L$ gauge bosons can be written in matrix form as
\begin{equation}
  \mathbf{A}_\mu = \frac{g_L}{\sqrt{2}}
  \begin{pmatrix}
    \frac{W^3_\mu}{\sqrt{2}}+\frac{W^8_\mu}{\sqrt{6}} &
    W^+_\mu & Y^{--}_\mu \\[4pt]
    W^-_\mu &
    -\frac{W^3_\mu}{\sqrt{2}}+\frac{W^8_\mu}{\sqrt{6}} &
    Y^{-}_\mu \\[4pt]
    Y^{++}_\mu & Y^{+}_\mu & -\frac{2W^8_\mu}{\sqrt{6}}
  \end{pmatrix}\!.
  \label{eq:Amu}
\end{equation}
The model predicts five new vector bosons: a  neutral boson $Z'$ and two bilepton doublets $(Y^{--},Y^-)$ and $(Y^{++},Y^+)$ carrying lepton numbers $L=+2$ and $L=-2$, respectively.

The bilepton interactions with fermions come from the
$SU(3)_L$-covariant kinetic terms
$\mathcal{L}\supset i\bar{\Psi}\gamma^\mu D_\mu\Psi$.
For the first two quark families, which transform as triplets, the
coupling to the doubly-charged bilepton reads
\begin{equation}
  \mathcal{L}_{Yqq} =
  \frac{g_L}{\sqrt{2}}
  \Bigl(
    \bar{u}_L\gamma^\mu D_L\,Y^{--}_\mu
   +\bar{c}_L\gamma^\mu S_L\,Y^{--}_\mu
  \Bigr) + \mathrm{h.c.},
  \label{eq:Yqq12}
\end{equation}
while the third family, transforming as an anti-triplet, gives
\begin{equation}
  \mathcal{L}_{Yqq}^{(3)} =
  \frac{g_L}{\sqrt{2}}\,
  \bar{T}_L\gamma^\mu b_L\,Y^{--}_\mu + \mathrm{h.c.}\,.
  \label{eq:Yqq3}
\end{equation}
 These vertices lead to the $t$-channel production of bi-leptons $q\bar{q}\to Y^{++}Y^{--}$, which decay via the interaction
\begin{equation}
  \mathcal{L}_{Y\ell\ell} =
  \frac{g_L}{\sqrt{2}}
  \sum_{f=e,\mu,\tau}
\bar{\ell}^{\,c}_{fR}\,\gamma^\mu\,\ell_{fL}\,Y^{--}_\mu
  + \mathrm{h.c.},
  \label{eq:Yll}
\end{equation}
to two pairs of same-sign leptons.

The non-Abelian $SU(3)_L$ structure further generates cubic gauge vertices, including $gY^{++}Y^{--}$ and $Z^{\prime}Y^{++}Y^{--}$ couplings that contribute to $s$-channel production. After electroweak symmetry breaking, the SM-like Higgs couples to bilepton pairs through $(D_\mu\rho)^\dagger(D^\mu\rho)$,
\begin{equation}
  \mathcal{L}_{hYY} =
  g_L^2\,v_\rho\,h\,Y^{++}_\mu Y^{--\,\mu}
  +\frac{g_L^2}{2}\,h^2\,Y^{++}_\mu Y^{--\,\mu},
  \label{eq:hYY}
\end{equation}
enabling the Higgs-mediated production channel $q\bar{q}(gg)\to h\to Y^{++}Y^{--}$.

So far, we studied unbroken $SU(2)_L$, such that the fermions are weak eigenstates. After EW breaking, CKM elements will enter these vertices and lead to interesting effects in flavour physics~\cite{Buras:2013dea,Rodriguez:2004mw,CarcamoHernandez:2018iel,Descotes-Genon:2017ptp}. However, for LHC processes, these small off-diagonal couplings can be neglected. Furthermore, as we will show later, for the parameter space under consideration, all such bounds can be respected.

Concerning the breaking of $SU(3)_L\times U(1)_X\to SU(2)_L\times U(1)_Y $, different Higgs representations can be used. Nonetheless, the LHC phenomenology of bilepton is, to a good approximation, independent of the Higgs sector, as long as the new scalars are heavy enough not to impact the bilepton branching ratios significantly. To be specific, we choose a scalar sector containing three $SU(3)_L$ triplets,
$\rho\in(1,3,+1)$, $\eta\in(1,3,0)$, and $\chi\in(1,3,-1)$,
whose neutral components develop vacuum expectation values (vevs)
$v_\rho$, $v_\eta$, and $v_\chi$. Symmetry breaking proceeds in two
stages: first $SU(3)_L\times U(1)_X\to SU(2)_L\times U(1)_Y$,
driven by the large vev $v_\rho$, which generates masses for the TeV scale quarks
and gauge bosons, and subsequently
$SU(2)_L\times U(1)_Y\to U(1)_\text{em}$ through $v_\eta$ and
$v_\chi$, satisfying
\begin{equation}
  V \equiv \sqrt{v_\eta^2+v_\chi^2} = 246\,\text{GeV},
  \qquad
  \tan\beta = \frac{v_\eta}{v_\chi}.
  \label{eq:vevs}
\end{equation}
At tree-level, the charged gauge bosons masses are
\begin{equation}
  M_{Y^{\pm\pm}}^2 \simeq \frac{g_L^2}{2}\!\left(v_\rho^2+v_\eta^2\right),
  \quad
  M_{Y^{\pm}}^2   \simeq \frac{g_L^2}{2}\!\left(v_\rho^2+v_\chi^2\right),
  \label{eq:bileptonmass}
\end{equation}
so that $v_\rho\gg V$ naturally places the bileptons in the TeV range while reproducing the SM spectrum at low energies. Similarly, for order-one Yukawa coupling, also the new TeV-scale quarks have masses of order $v_\rho$. Furthermore, the vector-like quarks will decay with approximately the same branching ratio to $Y^\pm$ and $Y^{\pm\pm}$ and a SM quark, because both massive vectors have nearly the same mass. 

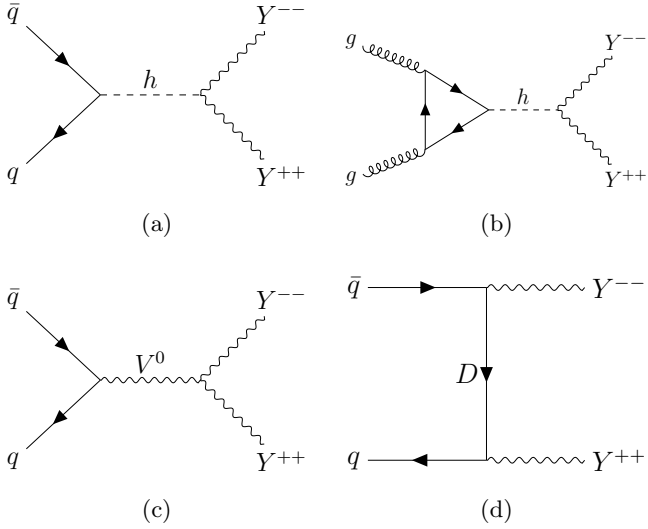
\begin{figure}[t]
\centering
\begin{minipage}[b]{0.48\columnwidth}
  \centering
  \resizebox{\linewidth}{!}{%
  \begin{tikzpicture}[every node/.style={font=\large}]
    \begin{feynman}
      \vertex (aqb) at (0,  1.3) {\(\bar{q}\)};
      \vertex (aq)  at (0, -1.3) {\(q\)};
      \vertex (av1) at (1.4,  0);
      \vertex (av2) at (3.0,  0);
      \vertex (aYm) at (4.3,  1.3) {\(Y^{--}\)};
      \vertex (aYp) at (4.3, -1.3) {\(Y^{++}\)};
      \diagram* {
        (aqb) -- [fermion]                   (av1),
        (aq)  -- [anti fermion]              (av1),
        (av1) -- [scalar, edge label=\(h\)]  (av2),
        (av2) -- [boson]                     (aYm),
        (av2) -- [boson]                     (aYp),
      };
    \end{feynman}
  \end{tikzpicture}}\\[4pt]
  (a)
\end{minipage}
\hfill
\begin{minipage}[b]{0.48\columnwidth}
  \centering
  \resizebox{\linewidth}{!}{%
  \begin{tikzpicture}[every node/.style={font=\large}]
    \begin{feynman}
      \vertex (bg1)  at (0,  1.3) {\(g\)};
      \vertex (bg2)  at (0, -1.3) {\(g\)};
      \vertex (blv1) at (1.4,  0.75);
      \vertex (blv2) at (1.4, -0.75);
      \vertex (blv3) at (2.6,  0);
      \vertex (bv2)  at (3.9,  0);
      \vertex (bYm)  at (5.2,  1.3) {\(Y^{--}\)};
      \vertex (bYp)  at (5.2, -1.3) {\(Y^{++}\)};
      \diagram* {
        (bg1)  -- [gluon]                        (blv1),
        (bg2)  -- [gluon]                        (blv2),
        (blv1) -- [fermion]                      (blv3),
        (blv3) -- [fermion]                      (blv2),
        (blv2) -- [fermion]                      (blv1),
        (blv3) -- [scalar, edge label=\(h\)]     (bv2),
        (bv2)  -- [boson]                        (bYm),
        (bv2)  -- [boson]                        (bYp),
      };
    \end{feynman}
  \end{tikzpicture}}\\[4pt]
  (b)
\end{minipage}

\vspace{1.2em}

\begin{minipage}[b]{0.48\columnwidth}
  \centering
  \resizebox{\linewidth}{!}{%
  \begin{tikzpicture}[every node/.style={font=\large}]
    \begin{feynman}
      \vertex (cqb) at (0,  1.3) {\(\bar{q}\)};
      \vertex (cq)  at (0, -1.3) {\(q\)};
      \vertex (cv1) at (1.4,  0);
      \vertex (cv2) at (3.0,  0);
      \vertex (cYm) at (4.3,  1.3) {\(Y^{--}\)};
      \vertex (cYp) at (4.3, -1.3) {\(Y^{++}\)};
      \diagram* {
        (cqb) -- [fermion]                         (cv1),
        (cq)  -- [anti fermion]                    (cv1),
        (cv1) -- [boson, edge label=\(V^0\)]       (cv2),
        (cv2) -- [boson]                           (cYm),
        (cv2) -- [boson]                           (cYp),
      };
    \end{feynman}
  \end{tikzpicture}}\\[4pt]
  (c)
\end{minipage}
\hfill
\begin{minipage}[b]{0.48\columnwidth}
  \centering
  \resizebox{\linewidth}{!}{%
  \begin{tikzpicture}[every node/.style={font=\large}]
    \begin{feynman}
      \vertex (dqb) at (0,  1.3) {\(\bar{q}\)};
      \vertex (dq)  at (0, -1.3) {\(q\)};
      \vertex (dv1) at (2.0,  1.3);
      \vertex (dv2) at (2.0, -1.3);
      \vertex (dYm) at (4.0,  1.3) {\(Y^{--}\)};
      \vertex (dYp) at (4.0, -1.3) {\(Y^{++}\)};
      \diagram* {
        (dqb) -- [fermion]                          (dv1),
        (dq)  -- [anti fermion]                     (dv2),
        (dv1) -- [fermion, edge label'=\(D\)]       (dv2),
        (dv1) -- [boson]                            (dYm),
        (dv2) -- [boson]                            (dYp),
      };
    \end{feynman}
  \end{tikzpicture}}\\[4pt]
  (d)
\end{minipage}
\caption{Representative Feynman diagrams for $Y^{++}Y^{--}$ production:
(a) $q\bar{q}$ annihilation via $s$-channel Higgs,
(b) gluon fusion via quark triangle loop,
(c) $q\bar{q}$ annihilation via $s$-channel gauge boson $V^{0}$,
(d) $q\bar{q}$ via $t$-channel exotic quark $Q$ exchange.}
\label{fig:feynman1}
\end{figure}

In this minimal structure, the mass relation between $Z^\prime$ and $Y^{\pm\pm}$ is \cite{Ng:1992st,RamirezBarreto:2008wq}
\begin{equation}
    \frac{M_{Y^{\pm\pm}}}{M_{Z^\prime}} \simeq \frac{\sqrt{3-12\sin^2\theta_w}}{2\cos\theta_w} \sim 0.3.
\end{equation}
The lower mass bounds on the $Z^\prime$ boson can be obtained by recasting the direct searches of ATLAS~\cite{ATLAS:2019erb} and CMS~\cite{CMS:2018ipm} for a sequential $Z^\prime$ boson and subsequently translating them into a corresponding lower bound on the bilepton mass.

Finally, in the bilepton model, the gauge couplings satisfy
\begin{equation}
    \frac{g_X^2}{g_L^2}
  = \frac{6\,\sin^2\theta_W(\mu)}{1 - 4\,\sin^2\theta_W(\mu)}\,,
\end{equation}
so that $g_X$ becomes indeterminate when $\sin^2\theta_W(\mu)\to \tfrac14$. Using the Standard Model running of $\sin^2\theta_W$, this occurs already at
$\mu \sim (3.8\text{--}4.0)\,\text{TeV}$, implying that the $SU(3)_L\times U(1)_X$ breaking scale cannot be above $\approx 4.0$\, TeV~\cite{Dias:2004dc}.

\section{Phenomenology}

In this section, we analyse bilepton pair production at the (HL) LHC, including both direct pair production and production via VLQ decays. The former signature consists of two same-sign lepton pairs with an invariant mass matching the bilepton mass. The latter includes, in addition, two jets. While these two jets are not relevant if $m_D>m_Y$, such that the bilepton is on-shell, it helps to discriminate signal from background for $m_D<m_Y$ where the bileptons are off-shell, and the invariant mass of the leptons is not peaking at $m_Y$.

\subsection{Setup}

For the phenomenological analysis, the model Lagrangian is implemented in \texttt{SARAH}~\cite{Staub:2015kfa}, which derives the full set of Feynman rules. The mass spectrum is then computed numerically with \texttt{SPheno}~\cite{Porod:2003um,Porod:2011nf}. For the LHC event simulation, we employ \texttt{MadGraph5\_aMC@NLO}~\cite{Alwall:2014hca,Frederix:2018nkq} for parton-level event generation, followed by \texttt{Pythia~8.3}~\cite{Bierlich:2022pfr} for parton showering and hadronization. Detector effects are included via \texttt{Delphes}~\cite{deFavereau:2013fsa}, employing the default ATLAS detector card.\footnote{Note that for our purpose, the differences between the ATLAS and the CMS detector are minor. The factorization and renormalization scales are set dynamically on an event-by-event basis using the default \texttt{MadGraph} prescription. Jets are reconstructed with the \texttt{FastJet} package~\cite{Cacciari:2011ma} using the anti-$k_T$ algorithm~\cite{Cacciari:2008gp} with a radius parameter $R=0.4$.} 

\begin{figure}[t]
    \includegraphics[width=\linewidth]{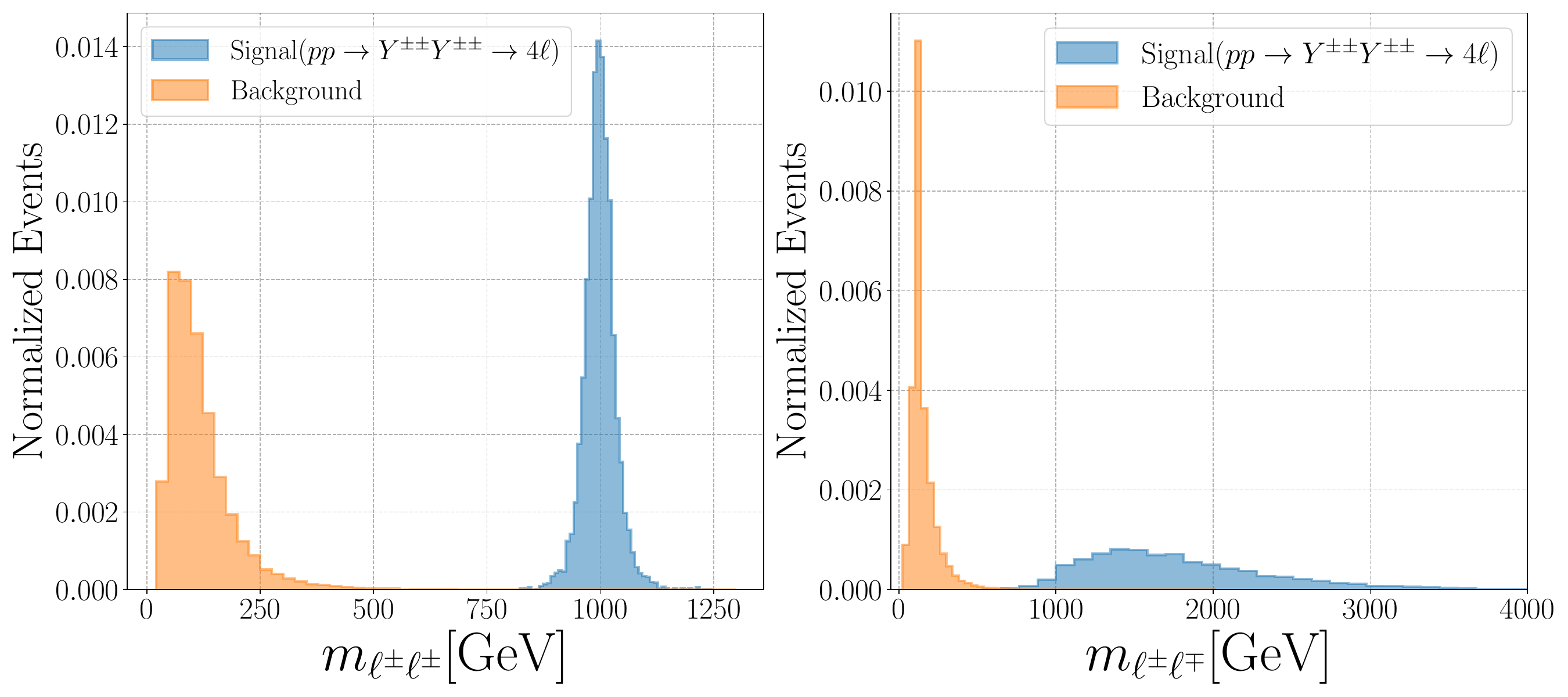}
    \caption{Invariant mass distributions of the same-sign lepton pair (left) and opposite-sign lepton pair (right), with $\ell = e , \mu$. Signal events are shown in blue, while background events are shown in orange. Each distribution is normalised to unity. The signal is generated for $m_{Y^{\pm\pm}} = 1\,\mathrm{TeV}$.}
    \label{fig:kin}
\end{figure}

Before turning to the collider analysis, let us briefly summarise the constraints from flavour and electroweak precision observables. The non-universal assignment of the third quark family under $SU(3)_L$ induces tree-level flavour-changing neutral currents (FCNCs) mediated by the $Z'$ boson when going from the weak eigenbasis to the physical basis with diagonal quark masses~\cite{Buras:2013dea,Rodriguez:2004mw}. The most stringent bound originates from $B_s$--$\bar{B}_s$ mixing, which can however typically be avoided for $m_{Z'}\gtrsim \text{3\,TeV}$~\cite{Buras:2013dea,Buras:2012dp,Descotes-Genon:2017ptp}. Limits of similar strength arise from direct searches for $Z'$ bosons at ATLAS~\cite{ATLAS:2019erb} and CMS~\cite{CMS:2018ipm} (once rescaled by the appropriate production cross section and branching ratios). Taking into account $M_{Y^{\pm\pm}}/M_{Z'} \simeq 0.3$, we have a lower bound on the bilepton mass of around 1\,TeV.\footnote{While the CMS result excludes a sequential-SM $Z'$ with masses below approximately $4.5\,\mathrm{TeV}$, this bound is relaxed in the 331 model due to the presence of additional decay channels of the $Z'$ into bileptons and VLQs. For relatively light bilepton and VLQ masses, the $Z'$ decays predominantly into these new states, thereby weakening the dilepton constraints. In contrast, for larger VLQ masses, the decay $Z' \to Q\bar{Q}$ becomes phase-space-suppressed, thereby enhancing the branching ratio into leptons and consequently strengthening the experimental bounds.} Electroweak precision observables, encoded in the oblique parameters $S$, $T$, and $U$, as well as constraints from $Z$-pole measurements and neutral current processes, impose additional bounds on the parameter space of the model~\cite{Ng:1992st} which are, however, weaker. 

In the following, we consider degenerate masses of the VLQs, $ m_{D_{1,2}} = m_T \equiv m_D$ (labelling the common mass by $m_D$) and fix $g_L$ through its relation to the weak coupling at the electroweak scale.

\subsection{Direct Bilepton Production}

The leading-order Feynman diagrams giving rise to direct bilepton production are shown in Fig.~\ref{fig:feynman1}. Followed by their decays to same-sign dilepton pairs, with $\ell = e,\mu$, leads to 4 high-energetic leptons, which is quasi background free. The dominant contribution to the production cross section arises from $s$-channel processes mediated by vector bosons, while the remaining channels are subleading.

\begin{figure}[t]
    \centering
\includegraphics[width=1.04\linewidth]{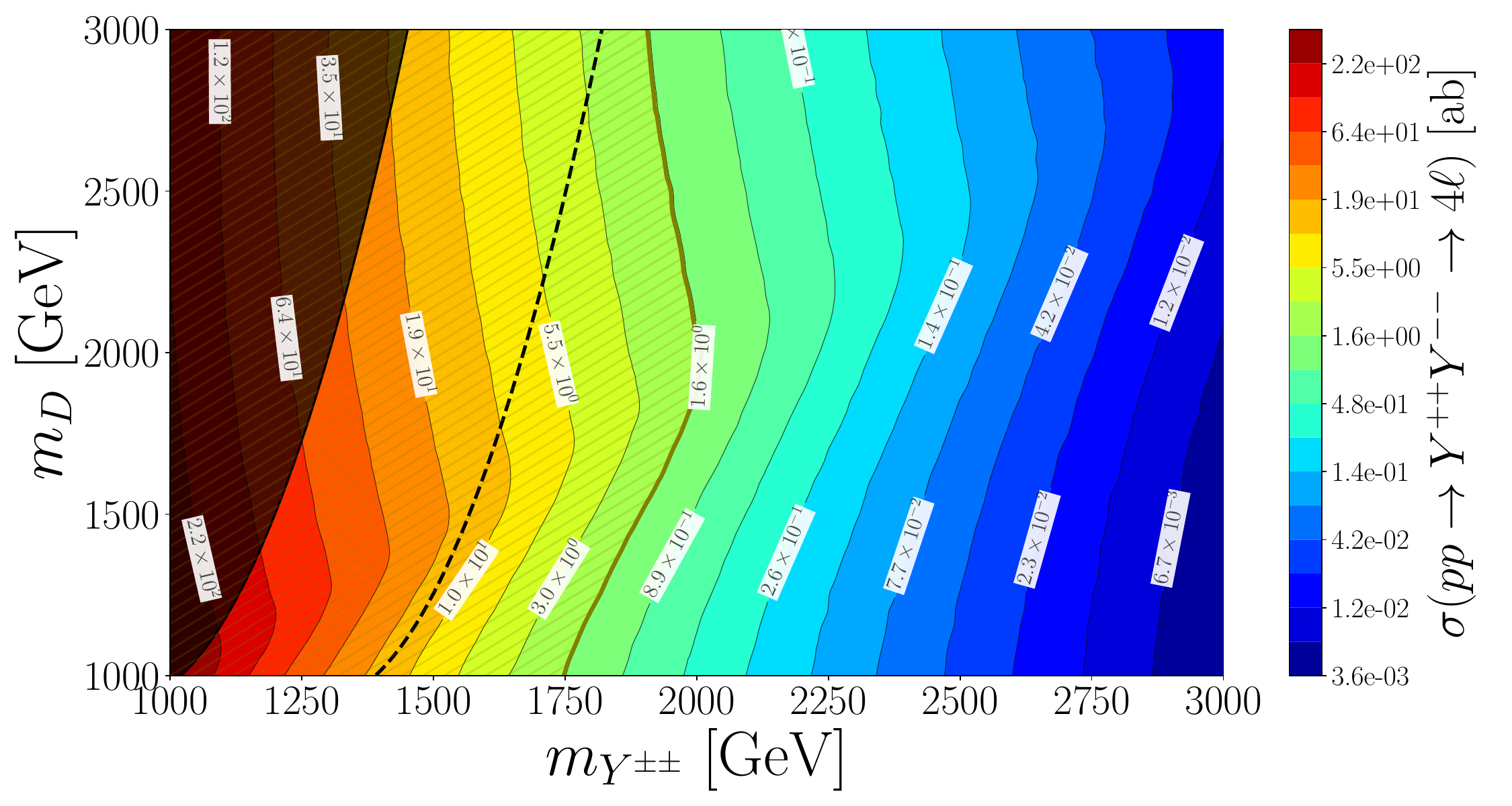}
    \caption{Visible cross section for $pp \to Y^{\pm\pm} Y^{\mp\mp} \to 4\ell$ in the bilepton–VLQ mass plane. The colour bar is given in units of attobarns. The dark-shaded region is excluded by the $Z^\prime \to \ell^+\ell^-$ search at the LHC, while the dashed black line denotes the projected exclusion at the HL-LHC with $\mathcal{L}=3000~\mathrm{fb}^{-1}$. The region with an expected number of events greater than five at $\mathcal{L}=3000~\mathrm{fb}^{-1}$ is shaded in olive.}
    \label{fig:scan1}
\end{figure}

To obtain a realistic estimate of the expected number of events, we impose the following detector-level selection cuts, ensuring particle isolation and efficient background suppression:
\begin{equation*}
    p_T(\ell) \ge 20\,\mathrm{GeV}, \hspace{4mm} \Delta R(\ell,\ell) \ge 0.1, \hspace{4mm} E_T^{\rm miss} \le 100\,\mathrm{GeV}\,,
\end{equation*}
where the requirement on $E_T^{\rm miss}$ is particularly effective in suppressing the $t\bar{t}Z$ background. After applying these cuts, the residual background is dominated by $ZZ$ processes. To further suppress it, one can consider the invariant mass distributions of the same-sign and opposite-sign lepton pairs,\footnote{To be more specific, the SM cross section for $pp\to ZZ\to 4\ell$ is around 24fb.  This would result in 150'000 events at the HL LHC. However, we generated one million events and no event survived the $m_{\ell\ell}>300$\,GeV cut. } as shown in Figure~\ref{fig:kin} for signal ($m_{Y^{\pm\pm}} = 1\,\mathrm{TeV}$) and background events. 

We now perform a scan in the $m_{Y^{\pm\pm}}$-$m_Q$ plane, as shown in Fig.~\ref{fig:scan1}.\footnote{The numerical scan in the two-dimensional 
$m_Y$-$m_Q$ plane presented below, is done with a uniform grid with 50\,GeV spacing in both directions. For each point, the relevant masses were updated in the input parameter card of \texttt{MadGraph}. The total decay widths of the bilepton and vector-like quarks were then computed at each scan point. Subsequently, the production cross section was calculated, and $10^5$ events were generated with the chain of packages discussed and the selection cuts were applied to determine the signal efficiency and thus the visible cross section. Finally, the grid points were interpolated.} The colour bar indicates the total cross section for the process $pp \to Y^{\pm\pm} Y^{\mp\mp} \to 4\ell$, expressed in attobarns, with $\ell = e,\mu$. The dependence on the masses can be understood as follows: For $m_{Y^{\pm\pm}} > m_Q$, the decay channel $Y\to Q\bar q$ (where $q$ is a SM quark) opens up such that the branching ratio to leptons gets suppressed. Furthermore, the cross section is suppressed kinematically by a rising $m_Y$ and for larger $m_Q$ the effect of diagram d) in Fig.~\ref{fig:feynman1} becomes smaller. This explains that the maximal cross section, for a given bilepton mass, occurs at $m_{Y^{\pm\pm}} = m_Q$. The black shaded region denotes the excluded parameter space from direct $Z'$ searches at the LHC, as discussed above. 

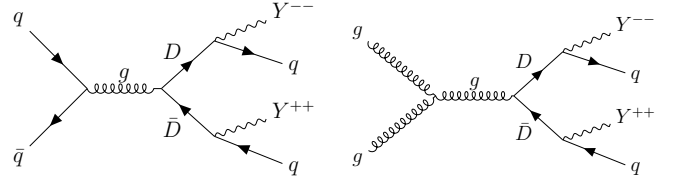
\begin{figure}[t]
\begin{minipage}[b]{0.48\columnwidth}
  \resizebox{\linewidth}{!}{%
  \begin{tikzpicture}[every node/.style={font=\large}]
    \begin{feynman}
      \vertex (eq)   at (0,  1.3) {\(q\)};
      \vertex (eqb)  at (0, -1.3) {\(\bar{q}\)};
      \vertex (ev1)  at (1.3,  0);
      \vertex (ev2)  at (2.7,  0);
      \vertex (ev3)  at (3.7,  0.9);
      \vertex (ev4)  at (3.7, -0.9);
      \vertex (eYm)  at (5.2,  1.5) {\(Y^{--}\)};
      \vertex (eqr1) at (5.2,  0.4) {\(q\)};
      \vertex (eYp)  at (5.2, -0.4) {\(Y^{++}\)};
      \vertex (eqr2) at (5.2, -1.5) {\(q\)};
      \diagram* {
        (eq)   -- [fermion]                                 (ev1),
        (eqb)  -- [anti fermion]                            (ev1),
        (ev1)  -- [gluon, edge label=\(g\)]                 (ev2),
        (ev2)  -- [fermion,      edge label=\(D\)]        (ev3),
        (ev2)  -- [anti fermion, edge label'=\(\bar{D}\)] (ev4),
        (ev3)  -- [boson]                                   (eYm),
        (ev3)  -- [fermion]                                 (eqr1),
        (ev4)  -- [boson]                                   (eYp),
        (ev4)  -- [anti fermion]                            (eqr2),
      };
    \end{feynman}
  \end{tikzpicture}}
\end{minipage}
\hfill
\begin{minipage}[b]{0.48\columnwidth}
  \resizebox{\linewidth}{!}{%
  \begin{tikzpicture}[every node/.style={font=\large}]
    \begin{feynman}
      \vertex (fg1)  at (0,  1.3) {\(g\)};
      \vertex (fg2)  at (0, -1.3) {\(g\)};
      \vertex (fv1)  at (1.6,  0);
      \vertex (fv2)  at (3.2,  0);
      \vertex (fv3)  at (4.2,  0.9);
      \vertex (fv4)  at (4.2, -0.9);
      \vertex (fYm)  at (5.7,  1.5) {\(Y^{--}\)};
      \vertex (fqr1) at (5.7,  0.4) {\(q\)};
      \vertex (fYp)  at (5.7, -0.4) {\(Y^{++}\)};
      \vertex (fqr2) at (5.7, -1.5) {\(q\)};
      \diagram* {
        (fg1)  -- [gluon]                                   (fv1),
        (fg2)  -- [gluon]                                   (fv1),
        (fv1)  -- [gluon, edge label=\(g\)]                 (fv2),
        (fv2)  -- [fermion,      edge label=\(D\)]        (fv3),
        (fv2)  -- [anti fermion, edge label'=\(\bar{D}\)] (fv4),
        (fv3)  -- [boson]                                   (fYm),
        (fv3)  -- [fermion]                                 (fqr1),
        (fv4)  -- [boson]                                   (fYp),
        (fv4)  -- [anti fermion]                            (fqr2),
      };
    \end{feynman}
  \end{tikzpicture}}
\end{minipage}
\caption{Representative Feynman diagrams for the VLQ pair production channel of bileptons. }
\label{fig:feynman2}
\end{figure}

\subsection{Bi-Lepton Production from VLQ Decays}

In addition to direct pair production, bileptons can originate at the LHC from the Drell-Yan production of vector-like quarks, which then decay to bileptons~\cite{Corcella:2021mdl,Corcella:2025ypl,Calabrese:2023ryr},
\begin{equation}
  pp \;\to\; D\bar{D} \;\to\; q\bar{q}\, Y^{++}Y^{--}\,.
\end{equation}
Because VLQ pair-production is governed by QCD, its cross-section is driven by the strong coupling $\alpha_s$ and gluon
PDFs, rather than electroweak couplings, which can make it substantially
larger than the direct electroweak channel \footnote{We have checked the contribution from $t$-channel $Y$-boson exchange and find it to be small due to the small photon PDF and even smaller $Z$- and $W$-boson PDFs.
}. The effective bilepton production rate in this channel is
\begin{align}
  \sigma(pp \to  \ell^\pm \ell^\pm \ell^\mp \ell^\mp + \mathrm{jets})
  &= \sigma(pp \to D\bar{D}) \nonumber\\
  &\quad \times \mathrm{Br}(D \to q\,Y^{\pm\pm(*)})^2,
  \label{eq:vlqxsec}
\end{align}
Here, we will consider both the on- and off-shell decays
\begin{equation}
  D_i \;\to\; u_i\, Y^{--},
  \qquad
  \bar{T} \;\to\; b\, Y^{--}\,,
  \label{eq:Ddecay}
\end{equation}
shown in Fig.~\ref{fig:feynman2}. Note that due to the absence of $W-W^\prime$ mixing (at tree-level), and the smallness of $VLQ$ mixing with SM quarks, the dominant decay mode of the VLQs, involving the vertex in Eq.~(\ref{eq:Yqq12}), still proceeds via bileptons even for $m_D<m_Y$. In our analysis, we include the contributions from all three VLQ species, assuming a common mass $m_D = m_T$, consistent with the setup described in the previous section.

\begin{figure}[t]
    \includegraphics[width=\linewidth]{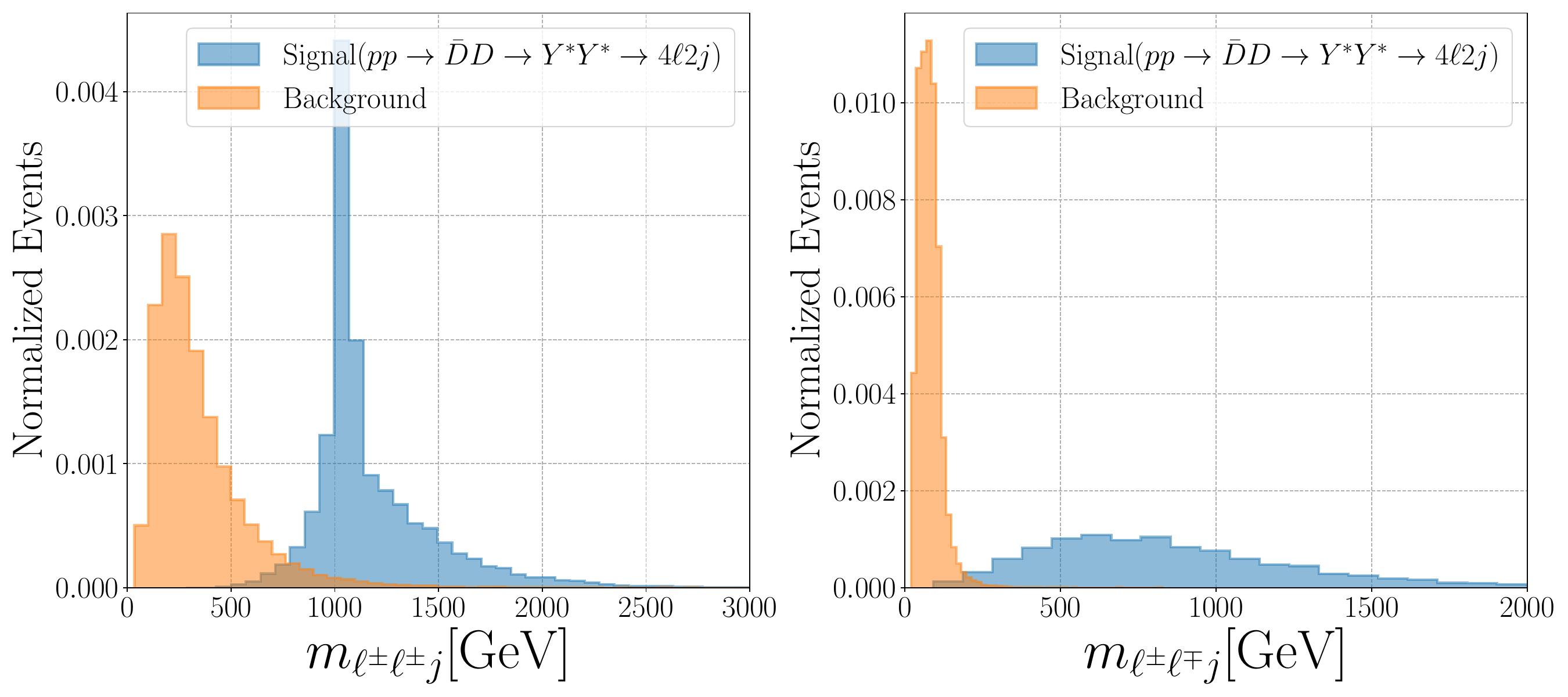}
    \caption{Invariant mass distributions of the same-sign lepton pair plus jet (left) and opposite-sign lepton pair plus jet (right), with $\ell = e, \mu$. Signal events are shown in blue, while background events are shown in orange. The signal is generated for  $m_{D} = 1\,\mathrm{TeV}$ and $m_{Y^{\pm\pm}} = 2\,\mathrm{TeV}$. The jet is paired such that the invariant mass is closest to the VLQ mass.}
    \label{fig:kin2}
\end{figure}

While in the regime where $m_D > m_Y$, bileptons can be produced on-shell, leading to a peak in the same-sign lepton pair invariant mass, resembling those in Fig.~\ref{fig:kin}, when $m_D < m_Y$, bileptons are produced off-shell, and such a peak is absent. Nevertheless, a distinct peak can still be observed around the vector-like quark mass in $\ell^{\pm}\ell^\pm j$ invariant mass distributions, as illustrated in Fig.~\ref{fig:kin2}. The dominant SM background for this process arises from $ZZ+\text{jets}$. However, no background events survive after applying a selection cut of $m_{\ell\ell}>600$ GeV (even at the HL LHC).

\begin{figure}[!ht]
    \centering
\includegraphics[width=1.02\linewidth]{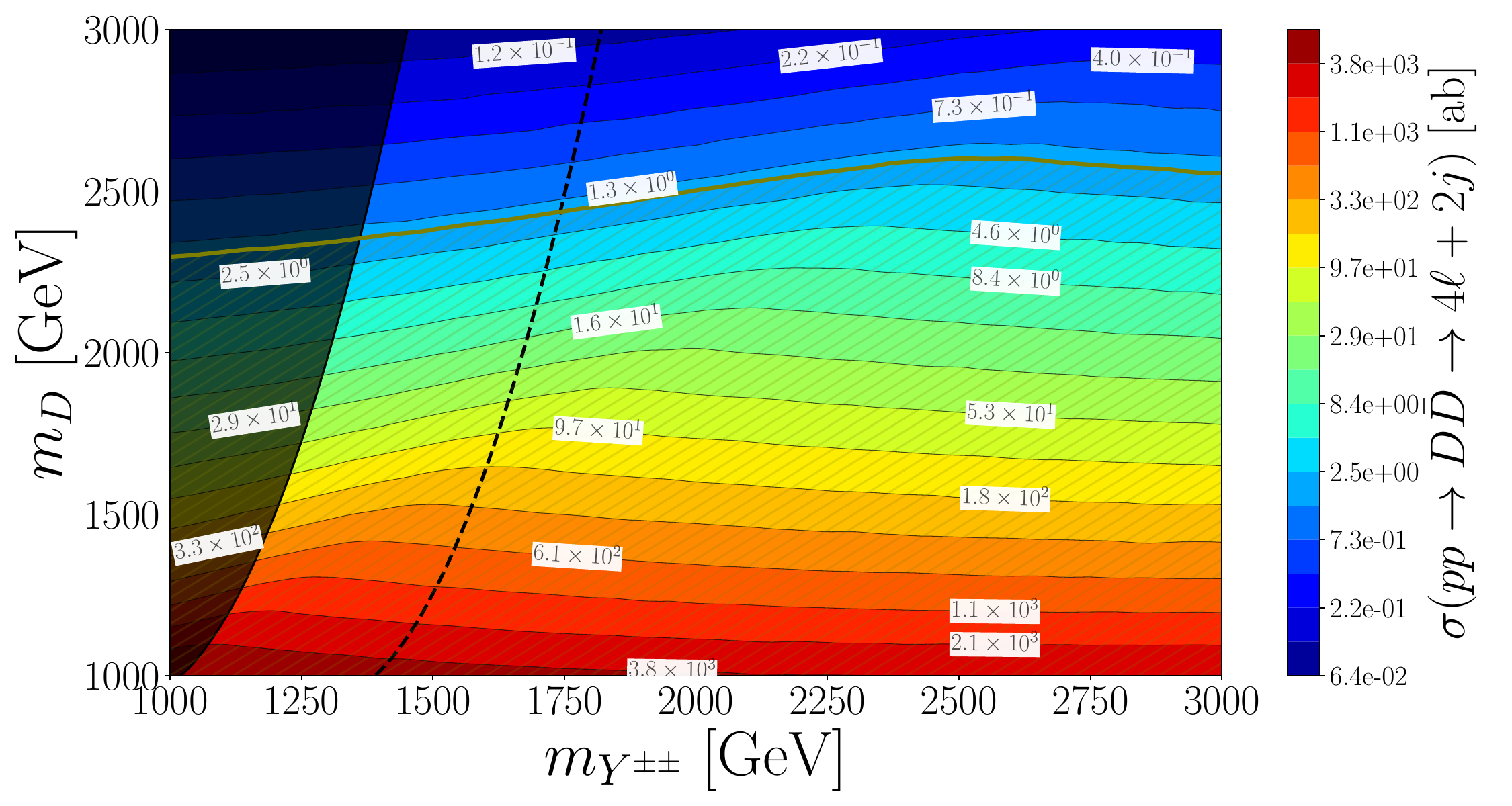}
    \caption{Visible cross section for $pp \to D\bar{D} \to Y^{++}Y^{--}
    + \mathrm{jets} \to 4\ell + \mathrm{jets}$ in the bilepton--VLQ
    mass plane, considering the on- and off-shell production of bileptons. The colour bar is given in units attobarns. The dark-shaded region is excluded by the $Z^\prime \to \ell^+\ell^-$ search at the LHC, while the dashed black line denotes the projected exclusion at the HL-LHC with $\mathcal{L}=3000~\mathrm{fb}^{-1}$. The region with an expected number of events greater than five at $\mathcal{L}=3000~\mathrm{fb}^{-1}$ is shaded in olive.}
    \label{fig:scan2}
\end{figure}

We perform the same scan over the bilepton--VLQ mass plane as in the previous subsection and present the resulting effective cross section for $pp \to 4\ell + \mathrm{jets}$ in Fig.~\ref{fig:scan2}. We adopt the same selection cuts used in the bilepton pair production analysis, with the additional requirement of the presence of two energetic jets satisfying an angular separation of $\Delta R(j,j) > 0.4$.  The qualitative behaviour differs significantly from the direct production case shown in figure~\ref{fig:scan1}: the cross sections are generally larger, leading to a broader coverage of the parameter space, and show a very weak dependence on $m_Y$.

\section{Conclusions and Implications}

\begin{figure*}[!hbt]
\includegraphics[width=0.95\linewidth]{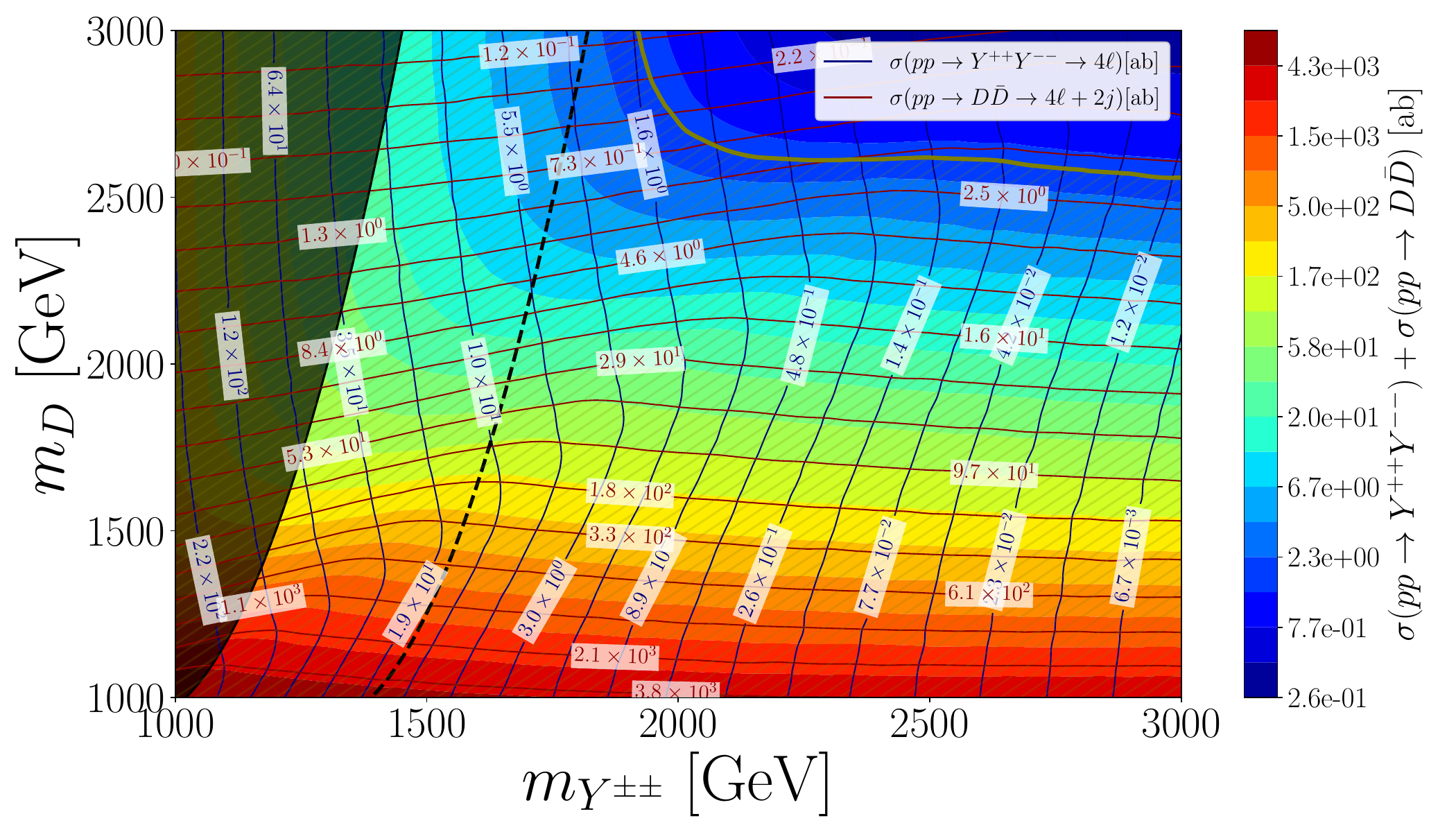}
    \caption{Visible cross section at a center-of-mass energy of 14\,TeV for bi-leptons. The blue contour lines represent the cross section for direct bilepton pair production, while red contour lines correspond to the cross section for vector-like quark pair production. The latter includes the off-shell production of bileptons. The colorbar indicates the sum of both cross sections, expressed in attobarns. The dark-shaded region is excluded by the $Z^\prime \to \ell^+\ell^-$ search at the LHC, while the dashed black line denotes the projected exclusion at the HL-LHC with $\mathcal{L}=3000~\mathrm{fb}^{-1}$. The region with an expected number of events greater than five at $\mathcal{L}=3000~\mathrm{fb}^{-1}$ is shaded in olive.}
    \label{fig:combined}
\end{figure*}

Extending the SM EW gauge symmetry to $SU(3)_L\times U(1)$ provides a natural explanation for the existence of three families of fermions. Furthermore, it treats the third quark generation differently, as expected from the hierarchy of fermion masses and CKM elements. A striking feature of this gauge group with the classic choice $\beta=\sqrt{3}$ is the presence of doubly-charged gauge bosons coupling to leptons, called bileptons. These particles have very distinct collider signatures that are quasi-free of SM background events. Furthermore, requiring the embedding of $SU(2)_L$ into $SU(3)_L$ places an upper limit on their masses of a few TeV, making them very promising candidates for a discovery at the LHC.

By simulating the bilepton pair production at the LHC and scanning over the $m_Y$--$m_D$ plane, we have shown why it is plausible that bileptons were not discovered with the LHC Run-2 data (140fb$^{-1}$) dataset, but can be observed at the HL LHC~\cite{Apollinari:2015bam,Apollinari:2017lan}: The region with a sufficiently high cross section for a Run-2 discovery is already largely excluded by direct searches for $Z^\prime$ bosons. On the other hand, the HL-LHC has the potential to discover bileptons with masses up to a few TeV, especially if the vector-like quarks are lighter than around 2\,TeV, as can be seen in Fig.~\ref{fig:combined}, where we combined the two production mechanisms for bi-leptons, including the off-shell case.

Furthermore, the discovery of the bilepton will imply the existence of three exotic quarks $(D, S, T)$ with TeV-scale masses as well as neutral and singly charged gauge bosons ($Z^\prime$ and $W^\prime$). This provides scientific motivation for building even larger colliders, such as the FCC-ee~\cite{FCC:2018evy} and FCC-hh~\cite{FCC:2018vvp}, where their effects on precision observables can be studied, and direct discoveries are possible, respectively.

There are also important theoretical consequences in case the bilepton model is confirmed. Because the three families are no longer identical, beginning at the TeV scale, the assumption of identical families made in $SU(5)$ grand unification~\cite{Georgi:1974sy} is no longer valid, and a similar objection arises to related attempts at QCD-electroweak unification based on higher rank groups like $SO(10)$ and $E(6)$. Although $SU(5)$ was widely believed during 1974-84, in retrospect, it would have been truly revolutionary had it agreed with proton decay because physics would have jumped by an unprecedented twelve orders of magnitude in mass and length scales. The bilepton model does not appear to be amenable to such unification with QCD and thus avoids proton decay. 

Furthermore, we may ask whether string theory can predict ${\cal N}_f$?  This can carry more weight as one author (PHF) has recently been credited~\cite{Nielsen:2024iln} with inventing string theory during the two weeks of September 9-23 of 1968, as the first to factorise the Veneziano model in terms of an infinite number of harmonic oscillators. A 1985 paper~\cite{Candelas:1985en} related ${\cal N}_f$ to the topological Euler number $\chi$ of the Calabi-Yau manifold used to compactify six spatial dimensions by  ${\cal N}_f = \frac{1}{2} |\chi|$. In 1987, however, another paper~\cite{Lerche:1986cx} revealed a very large number $\sim 10^{1500}$ of choices in the compactification. This  ambiguity remains in current discussions of the landscape and swampland, so rendering a string theory prediction
of ${\cal N}_f=3$, or any of the other 28 SM parameters, unlikely.

\section*{Acknowledgments}
A. Hammad is funded by grant number 22H05113, “Foundation of Machine Learning Physics”, Grant-in-Aid for Transformative Research Areas, and 22K03626, Grant-in-Aid for Scientific Research (C). A. Hammad is partially supported by the Science, Technology \& Innovation Funding Authority (STDF) under project ID 50806.

\bibliographystyle{JHEP}
\bibliography{bib.bib}

\end{document}